\documentclass[11pt]{article}
\usepackage{moriond,epsf}
\usepackage{times}

\newcommand{\epm}[2]{
 \raisebox{-0.5ex}{\shortstack[l]{$\scriptstyle+#1$\\$\scriptstyle-#2$}}}

\newcommand{\lt}{\left}
\newcommand{\rt}{\right}
\newcommand{\no}{\nonumber }                 
\newcommand{\ov}{\overline}
\newcommand{\eq}[1]{(\ref{#1})}

\newcommand{\prl}{Phys.~Rev. Lett.}
\newcommand{\prd}{Phys.~Rev.~D}

\newcommand{\npb}{Nucl.~Phys.~B}

\newlength{\nseparation}
\newenvironment{nfigure}[1]
        {\begin{figure}[#1]\hrule\vspace{\nseparation}\par}
        {\vspace{\nseparation}\par \hrule \end{figure}}

\begin{document}

\title{{\normalsize DESY-98-053 \hfill hep-ph/9805388} \\[2cm]
INCLUSIVE DIRECT CP-ASYMMETRIES IN CHARMLESS 
$\mathbf{B^{\pm}}$--DECAYS\footnote{Talk given at the \emph{Recontre de
    Moriond}\ on QCD and High Energy Hadronic Interactions, 
    March 21st to 28th, 1998, Les Arcs, France, to appear in the proceedings.}
}

\author{ Ulrich Nierste }
\address{DESY - Theory group, Notkestrasse 85, D-22607 Hamburg, Germany}
\maketitle\abstracts{
  Direct CP-asymmetries in inclusive decay modes can be 
  cleanly calculated by exploiting quark-hadron duality. This is in
  sharp contrast to CP-asymmetries in exclusive channels, where
  unknown strong phases prevent a clean extraction of CKM parameters 
  from measured CP-asymmetries. We have calculated the inclusive 
  CP-asymmetries in $B^{\pm}$--decays into charmless final states with 
  strangeness one or strangeness zero. In our results large logarithms
  are properly summed to all orders.  We find  
  \begin{eqnarray} 
    a_{CP}\lt( \Delta S=0 \rt) \; = \; \lt( 2.0 \epm{1.2}{1.0} \rt) \%,
      && \qquad 
    a_{CP}\lt( \Delta S=1 \rt) \; = \; \lt( -1.0 \pm 0.5 \rt) \% . \no
  \end{eqnarray}
  The constraints on the apex $(\ov{\rho},\ov{\eta})$ of the unitarity
  triangle obtained from these two CP-asymmetries define circles in
  the $(\ov{\rho},\ov{\eta})$-plane. $ a_{CP}\lt( \Delta S=0 \rt)$ 
  measures $\sin \gamma \cdot |V_{cb}/V_{ub}|$. The presented work has
  been done in collaboration with Gaby Ostermaier and Alexander Lenz.
 }
First we define the average branching fraction in terms of the decay
rate $\Gamma$:
\begin{eqnarray}
\ov{Br} &=& \frac{  \Gamma \lt( B^+ \rightarrow X \rt) + 
 \Gamma \lt( B^- \rightarrow \ov{X} \rt) }{2 \Gamma_{tot}} . \label{defbr}
\end{eqnarray}
In the following we are interested in inclusive final states $X$
containing no charmed particle. Further the total strangeness $S$ of $X$ 
must be known, we will consider the cases $X=X \lt( S=0 \rt)$ and 
$X=X \lt( \lt| S \rt| =1 \rt)$. 
Similarly we define the direct CP-asymmetries as
\begin{eqnarray}
A_{CP} &=& \frac{1}{2} \lt[ Br \lt( B^+ \rightarrow X \rt) - 
 Br \lt( B^- \rightarrow \ov{X} \rt) \rt], \qquad \quad
a_{CP} \; = \; \frac{A_{CP}}{\ov{Br} } . \label{defacp}
\end{eqnarray}
The measurement of $\ov{Br}$ in \eq{defbr} and of the CP-asymmetries
in \eq{defacp} requires a sum over semi-inclusive final states in
which the Kaons and strange baryons must be identified. Direct
CP-asymmetries in exclusive decay modes are hard to access
theoretically. Non-perturbative rescattering effects induce strong
phases, which are difficult to estimate. On the contrary for the case
of inclusive final states local quark-hadron duality allows to
calculate the quantities in \eq{defbr} and \eq{defacp} reliably within
perturbation theory. 
\begin{nfigure}{tb}
\vspace{-13mm}
\centerline{\epsfysize=0.5\textwidth \epsffile{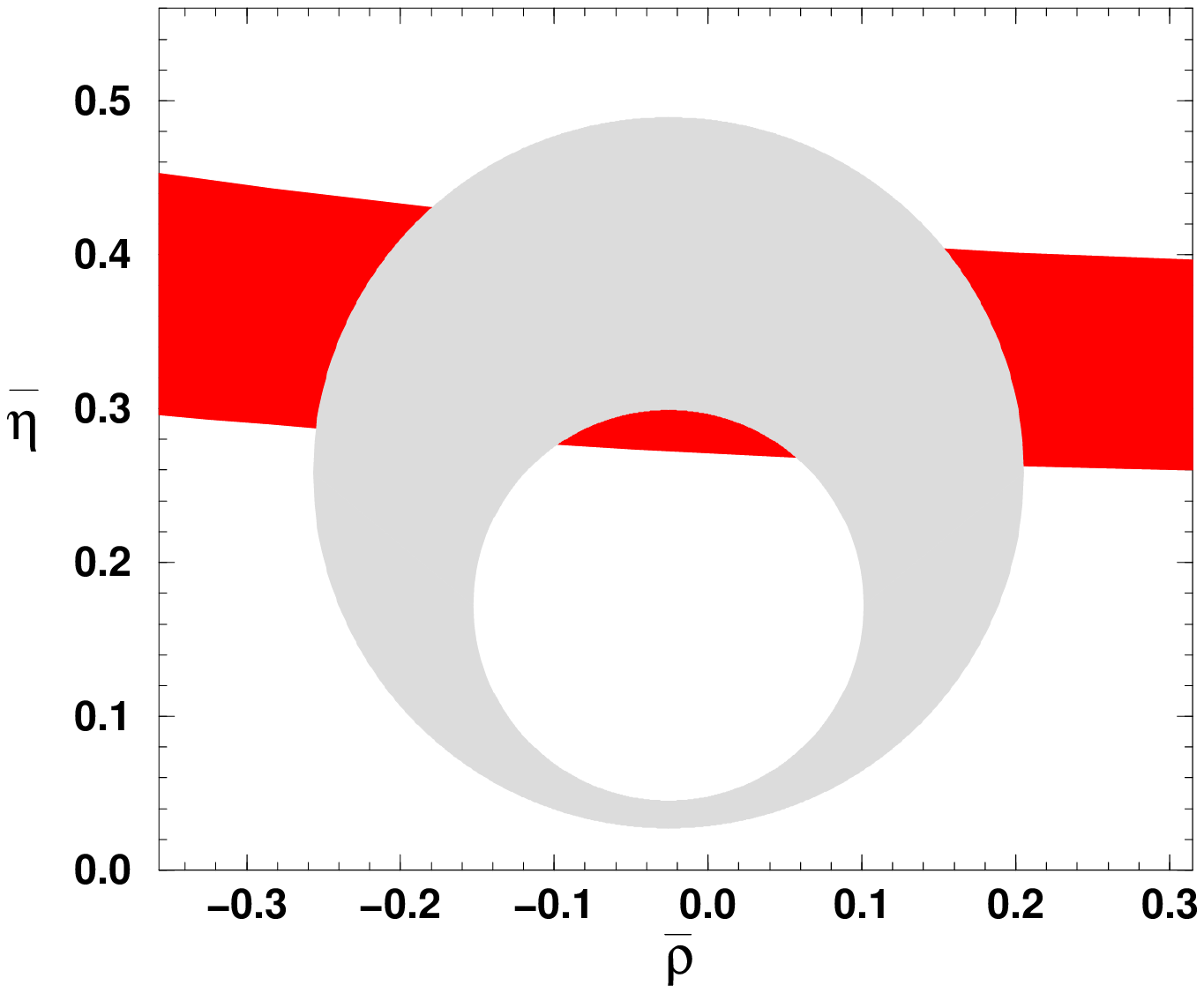}}
\vspace{-7mm}
\caption{The lightly (darkly) shaded area shows the constraint on 
  $(\ov{\rho},\ov{\eta})$ stemming from $a_{CP} \lt( \Delta S=0 \rt)$
  ($a_{CP} \lt( \Delta S=1 \rt)$), if the $a_{CP}$'s are measured as
  in \eq{exa}.  }\label{fig:acp}
\end{nfigure}
Theorist have been considering inclusive direct
CP-asymmetries since 1979 \cite{th}.  Up to now the inclusive
$a_{CP}$'s were believed to be small, of the order of a few permille.
Yet in the predictions of \cite{th} either large logarithms have not
been summed to all orders or $m_t$ was taken too small. 
This has been corrected for in our new paper\cite{lno1}, in which also
the predictions for charmless branching ratios calculated before \cite{lno2}
have been improved by incorporating new QCD corrections. We find 
\begin{eqnarray} 
    a_{CP}\lt( \Delta S=0 \rt) \; = \; \lt( 2.0 \epm{1.2}{1.0} \rt) \%,
      && \qquad 
    a_{CP}\lt( \Delta S=1 \rt) \; = \; \lt( -1.0 \pm 0.5 \rt) \% . \no
\end{eqnarray}
Here the error bars stem from the uncertainty in $m_c/m_b$ and 
$\ov{\rho},\ov{\eta}$ and from the residual dependence on the
renormalization scale $\mu$. The $\mu$-dependence can be reduced by 
calculating certain two-loop diagrams. This is possible with
reasonable effort and will be done, once the inclusive direct
CP-asymmetries receive experimental interest.  

The dependence of the $a_{CP}$'s on $\ov{\rho}$ and $\ov{\eta}$ is 
welcome in order to constrain the apex of the unitarity triangle. 
For a model scenario with 
\begin{eqnarray}
a_{CP} \lt( \Delta S=0 \rt) \; = \; 2.0 \, \% ,&& \qquad  \qquad  
a_{CP} \lt( \Delta S=1 \rt) \; = \; -1.0 \, \%  \label{exa}
\end{eqnarray}
and an assumed total error of $20 \, \% $ the constraints on
$\ov{\rho},\ov{\eta}$ are shown in figure \ref{fig:acp}. They
are nice circles in the $\ov{\rho},\ov{\eta}$-plane, whose
information is complementary to the familiar circle from
$B$-$\ov{B}$-mixing and the hyperbola from $\varepsilon_K$\cite{muc}.
The $A_{CP}$'s defined in \eq{defacp} are simply proportional to $\ov{\eta}$.

\end{document}